\documentclass{article}

\usepackage{graphicx}  
\usepackage{amsmath}   
\usepackage[compress]{cite}
\usepackage{amssymb}   
\usepackage{bm} 

\def\eq#1{{Eq.~(\ref{#1})}}

\def\zpl{{zero-point-length}}

\def\mdof{{microscopic degrees of freedom}}

\title{Gravitational effective action at mesoscopic scales from the quantum microstructure of spacetime}

\author{T. Padmanabhan\\
IUCAA, Post Bag 4, Ganeshkhind,
 Pune - 411 007, India.\\
email: paddy@iucaa.in}

\date{ }

\begin{document}

\maketitle

\begin{abstract}
At mesoscopic scales, the quantum corrected field equations of gravity  should arise from extremising, $\Omega$, the number of microscopic configurations of pre-geometric variables consistent with a given geometry. This $\Omega$, in turn, is  the product over all events $\mathcal{P}$ of the density, $\rho(\mathcal{P})$, of microscopic configurations associated with each event $\mathcal{P}$. One would have expected  $\rho\propto\sqrt{g}$ so that $\rho d^4x$ scales as the proper volume of a region.
 On the other hand, at leading order, we would expect the extremum principle to be based on the Hilbert action, suggesting $\ln\rho\propto R$. I show how these two apparently contradictory requirements can be reconciled by using the functional dependence of $\sqrt{g}$ on curvature, in the Riemann normal coordinates (RNC), and coarse-graining over Planck scales. This leads to the  density of microscopic configurations to be  $\rho = \Delta^{-1} = \sqrt{g}_{RNC}$ where $\Delta$ is the coarse grained Van-Vleck determinant. The approach also provides: (a) systematic way of computing QG corrections to field equations and (b) a direct link between the effective action for gravity and the kinetic theory of the spacetime fluid.  
\end{abstract}

A complete model for the  quantum spacetime is needed to probe \textit{microscopic} length scales $\lambda \lesssim L_P $ (where $L_P^2 = G\hbar/c^3$ is the Planck length) while the classical field equation, derived from the Hilbert action, is adequate for probing \textit{macroscopic}  length scales $\lambda \gg L_P$.  At the intermediate \textit{mesoscopic} scales  which are large --- but not significantly large --- compared to $L_P$, one would expect the spacetime to be described  in terms of an effective geometry, determined by the  field equations which incorporate the quantum gravity (QG) corrections. Can we address the dynamics of this regime  in terms of a suitable extremum principle which, at the leading order, will reduce to that based on the Hilbert action? I will show how this can be done, by combining classical geometric considerations  with coarse-graining over sub-Planckian degrees of freedom. 

An analogy between spacetime dynamics and fluid dynamics will be helpful to set the stage. In the study of a fluid made of discrete atoms (``\mdof'') one can delineate three levels of description. The first one is  \textit{microscopic} in which  one would describe the system in a completely quantum mechanical language by, say, writing the Schrodinger equation for all the atoms of the fluid. At the other extreme, we have a \textit{macroscopic} description, say, in terms of continuity and Navier-Stokes equations, which totally ignores the granularity of the fluid and treats it as a continuum. In this regime we would, for example, describe the velocity of the fluid at a given event $x^i$ by a single-valued vector function $\bm{v}(t, \bm{x})$. These two descriptions are analogous to the microscopic and macroscopic descriptions of the spacetime mentioned above. What we are looking for is a mesoscopic scale description interpolating between these two. The kinetic theory of fluids, in terms of a distribution function $f(x^i, p_i)$, provides such an interpolation between these two extremes. The distribution function recognizes the discreteness of the fluids --- with $dN = f(x^i, p_i)\, d\Gamma$ counting the \textit{number} of atoms in a phase volume $d\Gamma \equiv d^3\bm{x} \, d^3 \bm{p}$ --- and, at the same time, allows a continuum description at scales sufficiently larger than  the mean free path.  In this mesoscopic scale description we recognize the existence of velocity dispersion at any given event  $x^i$ due to  discrete atoms with different momenta $p_i$ co-existing at a given event. What we are interested in is a similar description, at mesoscopic scales, for the spacetime fluid with Planck length playing the role of the mean free path. 

To obtain such a description, we can proceed as follows: Consider a spacetime geometry $\mathcal{G}$, described a metric tensor $\bar g_{ab} ( \bar x)$ in an arbitrary  coordinate chart $\bar x^i$ in some region of spacetime. There will a large number, $\Omega$, of microscopic configurations of (as yet unknown) pre-geometric variables which will be consistent with any given emergent geometry $\mathcal{G}$. (This is similar to the existence of several microscopic configurations of a the atoms of a fluid, consistent with some macroscopic parameters like pressure, density etc). Let $f(\mathcal{P}, \xi_A)$ be  the number of possible configurations of the \mdof\ of spacetime, associated with an event $\mathcal{P}$, in a given geometry  $\mathcal{G}$. As indicated, this function $f(\mathcal{P}, \xi_A)$ could also depend on the relics of some sub-Planckian  degrees of freedom symbolically denoted by $\xi_A$. (These are analogous to the momenta of individual atoms  in kinetic theory of a fluid.)
 The total number of microscopic configurations in a given region of spacetime is then given by
\begin{equation}
 \Omega \equiv \prod_{\mathcal{P},\xi_A} f(\mathcal{P},\xi_A) = \exp [\sum_{\mathcal{P},\xi_A} \ln f(\mathcal{P},\xi_A)] \Rightarrow \exp\left [-\int {d^4\bar x} \ \sqrt{\bar g} \, \mathcal{L}_E\right]\equiv \exp -A_{E}
 \label{basic1}
\end{equation} 
where
\begin{equation}
\mathcal{L}_E=-L^{-4}\sum_{\xi_A}\ln f(\mathcal{P},\xi_A)\equiv -L^{-4}\ln \rho(\mathcal{P});\qquad
\rho(\mathcal{P})\equiv \prod_{\xi_A} f(\mathcal{P},\xi_A)
\end{equation}
Here $\rho(\mathcal{P})$ denotes the number of microscopic configurations associated with a given event $\mathcal{P}$ once the product over internal variables $\xi_A$ is taken.
In the third step in \eq{basic1}, we have taken the continuum limit and introduced the 
length-scale $L = \mathcal{O}(1) L_P \equiv \mu L_P$, of the order of Planck length,   to ensure proper dimensions. (That is, the discrete sum over events $\mathcal{P}$ goes over to integration with dimensionless proper volume measure $\sqrt{\bar g}d^4x/L^4$ in the continuum limit.) The $A_E$ can be thought of as an effective (Euclidean) action.
 The QG corrected field equations at mesoscopic scales can then be obtained by extremising the microscopic configurations $\Omega=e^{-A_E}$ with  $\mathcal{L}_E$  playing the role of a gravitational effective Lagrangian.
Our next task is to determine $f(\mathcal{P}, \xi_A)$ and $\rho$ from physical considerations.

 As I will show later on, the variables $\xi_A$ arise very naturally in this approach. However, let me ignore these  variables $\xi_A$ just for the moment so that $f$ is the same as $\rho$.
Then I would expect  $f=\rho \propto  \sqrt{\bar g}$ so that $\rho(\bar x)d^4\bar x\propto \sqrt{\bar g}(\bar x)d^4\bar x$ will scale as the proper volume of a region. The QG corrected field equations of the gravitational sector has to then come from the effective Lagrangian $\mathcal{L}_E = -L^{-4}\ln \sqrt{\bar g}$ when we vary the metric. 
At first sight, such an interpretation seems impossible because of three issues: (i) We would expect the extremum principle to be based on the Ricci scalar $R$ at the appropriate limit rather than on $\ln\sqrt{\bar g}$. (ii) One would like to interpret $\sqrt{\bar g}$ in terms of a suitable scalar quantity, after eliminating the gauge  freedom in the metric tensor at any event $\mathcal{P}$. (iii) One cannot expect to get a finite number of \mdof\ without invoking some QG considerations. I will now show how all these issues can be tackled with the proper interpretation of the $ \sqrt{\bar g}$ factor. 

The metric tensor $\bar g_{ab}(\bar x)$ (and  its determinant), at any given event $\mathcal{P}'$,   contain come amount of gauge redundancy 
in any generic coordinate system. So our first task is to eliminate this freedom at an event by choosing an appropriate coordinate system. Such a coordinate system, usually called Riemann normal coordinates, is well known in literature (see e.g.,\cite{rnc1,rnc2}). If one introduce the Riemann normal coordinates (RNC), centered around an event $\mathcal{P}'$ in the spacetime, the new metric $g_{ik}(\mathcal{P}, \mathcal{P}')$ at any another event $\mathcal{P}$ will depend on \textit{both} $\mathcal{P}$ and $\mathcal{P}'$. That is, in  the RN coordinates, we get a \textit{family} of metrics $g_{ik}(x,x')$ at  $x$, parameterized by the coordinates of another event $x'$. The construction of RNC ensures that metric reduces to the Cartesian form, and the Christoffel symbols vanish, at $x'$.

This choice also allows a co-ordinate invariant description. It is well-known that \cite{poisson}, the $\sqrt{g}(x,x')$ in RNC is precisely equal  to the reciprocal of the Van-Vleck\footnote{The Van-Vleck determinant is defined, in \textit{arbitrary} coordinates, by: $\Delta(x,x')=D(x,x')/(\sqrt{g(x)}\sqrt{g(x')})$ where $D=(1/2){\rm Det}[\partial_a\partial'_b\sigma^2(x,x')]$} determinant, 
$\Delta^{-1}(x,x')$, obtained from the second derivatives of the geodesic interval, $\sigma^2(x,x')$. That is, $\Delta^{-1}(x,x')$ is a (bi)scalar which reduces to $\sqrt{g}$ in RNC; so by using $\Delta^{-1}(x,x')$ in arbitrary coordinates, we also obtain a covariant description.\footnote{This trick of using $\Delta^{-1}$ to replace $\sqrt{g}$ in RNC in order to obtain a covariant description is well known in literature. See, e.g., \cite{parker1,parker2}.} 
The metric determinant $\sqrt{g}=\Delta^{-1}$ in RNC can be expressed in terms of the geometrical variables built from the curvature tensor $R_{ijkl}$ and its derivatives. It will also depend on the geodesic distance $\sigma^2(x,x')$ and the unit-norm vectors $n_i(x,x') \equiv \partial_i \sigma$ with the latter two appearing (only) through the combination $q^i(x,x') \equiv \sigma n^i$.
 We can, therefore, write down the functional 
\begin{equation}
\Delta^{-1}(x,x')= (\sqrt{g})_{RNC} = f[R_{ijkl}(x); \sigma n^i]=f[x; q^i(x,x')]
\end{equation}
  where it is understood that the functional dependence on the curvature also includes dependence on its derivatives.\footnote{One could have introduced a proportionality constant $\lambda$ between $(\sqrt{g})_{RNC}$ and $f$ and written $f=\lambda\sqrt{g}$. \textit{Observations} show that $\lambda=1$ to a very high order of accuracy! I will discuss this later. The choice $\lambda=1$ is equivalent to the normalisation $\rho=1$ in flat spacetime.} 
  
  This expression is non-local (i.e., it depends on two events $x$ and $x'$) while we are looking for a local expression for the density of \mdof\ associated with an event. The simplest procedure to get a local expression will be to take the limit $x\to x'$; this, however,  does not work because --- in this limit --- $\sigma(x,x)$ vanishes and all the dependence on the curvature disappears. This, of course, is to be expected. It will be impossible to obtain a non-trivial, local, measure of microscopic degrees of freedom without introducing some kind of `discreteness' in the spacetime at Planck scales. A point of strictly zero size cannot host degrees of freedom in spacetime, just as it cannot host a finite number of atoms in a fluid. In a fluid we need to average over a volume which is large compared to the mean free path; similarly, in the spacetime we need to coarse-grain over Planck-size regions.
  Further, $f[x,\sigma n^i]$  depends on the unit normalized vector field $n^i$ which needs to be properly interpreted. These two issues require making a reasonable postulate about microscopic physics, which I will now bring in. 
  
  There is significant evidence to suggest that the key relic of trans-Planckian physics, which survives at the mesoscopic scales, is the modification of the geodesic distance $\sigma^2$ by the addition of a \zpl.\footnote{There is extensive literature on this idea; for a sample, see \cite{zpl1,zpl2,zpl3,zpl4,zpl5,zpl6,zpl7,zpl8,zpl9,zpl10,zpl11,zpl12,zpl13,zpl14,zpl15,zpl16}. A succinct summary of the motivation and justification for the \zpl\ is given in ref. \cite{zplsummary1,zplsummary2}. As I will show later, the final result in this work  can also be obtained  by an alternative procedure \textit{without} introducing the \zpl. However, this is my preferred approach.}  That is, we postulate that the coarse-graining over microscopic scales will lead to the replacement: 
 $
 \sigma^2(x,x') \to \sigma^2_{QG}(x,x')\equiv \sigma^2(x,x') + L^2                               
 $ 
 where $L=\mu L_P=\mathcal{O}(1)L_P$ is the \zpl\ of the spacetime.  We can now take the local limit $x' \to x$ which leads to the replacement of $\sigma(x,x')$ in all the expressions by $\sigma_{\rm QG} (x,x) = L$.  Physically, this is equivalent to setting the geodesic distance between the events $x,x'$ to the \zpl\ --- which is a natural replacement for the coincidence limit --- when QG effects are taken into account. 
 
 We can now identify the density of \mdof\  as  $f[R_{ijkl}(x); Ln^i]=\Delta^{-1}[R_{ijkl}(x); Ln^i]$, with $n^i$ playing the role of internal variables $\xi_A$ in \eq{basic1}. 
 To determine the effective Lagrangian, $\mathcal{L}_E\propto \ln\rho$ at an event, we sum over all $n_i$ in $\ln f[R_{ijkl}(x); Ln^i]$,  preserving the unit norm ($n^2=1$).  We thus get the final expression for the Euclidean effective Lagrangian:
\begin{equation}
 \mathcal{L}_E=-L^{-4}\ln\rho[R_{ijkl}(x)]=L^{-4} \left[\int_{S^3} d^3n \ln \Delta[R_{ijkl}(x); Ln^i]\right]
 \label{effL}
\end{equation} 
Note that a natural set of internal degrees of freedom arises in the form of unit-norm vector field $n^i$ and are summed over. This $\mathcal{L}_E$ is a non-perturbative expression for the gravitational effective action at mesoscopic scales. It stresses the importance of geodesic interval $\sigma^2$ and the Van-Vleck determinant obtained from it. 

Since we do not have a closed expression for $\sqrt{g} = \Delta^{-1}$ in RNC, the effective Lagrangian in \eq{effL} has to be computed as a series in $L_P^2$. This is easily done using the known \cite{ricci,leo} series expansions, for, say, ${g}$ in RNC
\begin{equation}
g=1-\sigma^2 A_{ij}(n^in^j) -\sigma^3 A_{ijk} (n^in^jn^k) -\sigma^4 A_{ijkl}(n^in^jn^kn^l) +\mathcal{O}(\sigma^5)
\label{gexp}
\end{equation} 
where $A_{ij}=(1/3)R_{ij}, A_{ijk}=(1/6)\nabla_kR_{ij}$ and $A_{ijkl}=(1/180)[9\nabla_l\nabla_k R_{ij}+2R^p_{\phantom pijq}R_{pkl}^{\phantom{pkl}q}-10R_{ij}R_{kl}]$.
Computing $(-\ln\sqrt{g})$, correct to $\mathcal{O}(\sigma^6)$ and integrating over all $n^i$ on unit 3-sphere, we get the effective Lagrangian in \eq{effL} to be:
 \begin{equation}
  \mathcal{L}_E=L^{-4} \left[\int d^3n \ln \Delta[R_{ijkl}(x); Ln^i]\right]=\frac{\pi^2}{12L^2}\left[R+L^2Q + \mathcal{O} (L^4)\right] 
  \end{equation}
  where $L=\mu L_P$ and
  \begin{equation}
  Q\equiv \frac{1}{2}\left[\frac{1}{20}\square R+\frac{1}{90}R_{ab}R^{ab} +  \frac{1}{10}\nabla_a\nabla_b R^{ab}
  +\frac{1}{60}R_{abcd}R^{abcd}\right]
  \label{defQ}
 \end{equation}
 The coefficient of $R$ has to be fixed by comparing with the Newtonian limit, in terms of $G_N$ by setting the coefficient to $(16\pi L_P^2)^{-1}$; this leads to $\mu^2=4\pi^3/3$. So our final result for the effective Lagrangian, correct to $\mathcal{O} (L_P^4)$, is:
 \begin{equation}
  \mathcal{L}_E = \frac{1}{16\pi L_P^2}\left(R + \frac{4\pi ^3}{3} L_P^2 Q + \mathcal{O} (L_P^4)\right)
  \label{final1}
 \end{equation} 
 These expressions allow us to compute quantum corrections to Einstein's equations in a systematic manner.\footnote{The expansion in \eq{gexp} and hence the form of $Q$ (in \eq{defQ}) are dimension-independent. The relative numerical factor between $R$ and $Q$ term is dimension- dependent because integrals over $n^an^b$ and $n^an^bn^cn^d$  are dimension-dependent. The coefficient of $Q$ in \eq{final1} gets multiplied by $6/(d+2)$ in d-dimensions.} 
 \textit{The conceptual simplicity of the approach which has led to such a tangible result is note worthy.}
 
Let me now comment on several aspects of this result and the  possible future directions of research. Let me begin with some technical comments.

The effective Lagrangian in \eq{final1} has no cosmological constant term because we took $f=\sqrt{g}$. If we had taken, instead,  $f=\lambda\sqrt{g}$ with a proportionality constant $\lambda$, then the $\ln f$ in \eq{effL} would have contributed an extra $\ln\lambda$ terms appearing as a dimensionless cosmological constant $\Lambda L^2\simeq\ln \lambda$. This tells you that $\lambda=1+\mathcal{O}(10^{-123})$. \textit{That is, the natural choice $\lambda=1$ indeed predicts zero cosmological constant.} It also suggests that the observed cosmological constant in the universe is a non-perturbative relic \cite{tphp} from QG which changes $\lambda$ by a tiny amount from unity. 

In our approach, we started with a function $\ln \Delta(x^i, \sigma n^i)$, set $\sigma$ equal to the \zpl, $L$, and summed over $n^i$ on an \textit{unit 3-sphere}. Instead, we could have averaged (coarse-grained) $\ln \Delta(x^i, \sigma n^i)$ over the sub-Planckian 4-sphere by integrating the displacement $q^i\equiv \sigma n^i$ over a \textit{4-sphere} of Planck scale radius $L_0$, say. That is, we integrate $n^i$ over all directions  and integrate $\sigma$ in the range $(0,L_0)$. This procedure will lead to the same structure for the correction term $Q$ and makes clear that we are effectively integrating out sub-Planckian scales. In other words, the introduction of \zpl\ can be replaced by the operation of averaging over Planck scale spherical regions. So, if you are uncomfortable with the idea of \zpl, you can get the same structure for the effective Lagrangian by explicit coarse-graining over sub-Planckian regions.

The RNC is constructed so that it is locally inertial at a given event eliminating gauge degrees of the metric \textit{at that event}. A synchronous reference frame with line element $ds^2=d\sigma^2+\sigma^2 \gamma_{\alpha\beta}(\sigma,x^\alpha)dx^\alpha dx^\beta$, on the other hand, imposes the gauge conditions $g_{00}=1,g_{0\alpha}=0$ in a local \textit{region} and serves the same purpose. It uses the geodesic distance $\sigma$ as a ``radial'' coordinate. In this frame $\Delta^{-1}=\rho=(\gamma/\gamma_{flat})^{1/2}$, where $\sqrt{\gamma}$ is the unit area element of $\sigma=$ constant surfaces \cite{comment1}.  The density of \mdof\ can now be thought of as being proportional to the dimensionless  \textit{area measure} $\sqrt{\gamma}$.   The spatial surfaces $\sigma$= constant maps to equi-geodesic surfaces $\sigma(x,x')$ = constant in an \textit{arbitrary} frame.\footnote{In fact, the relation $f=\Delta^{-1}$ was first introduced in \cite{zpl13} (see eq 57) and was explored further
in the context of area of equi-geodesic surfaces in several previous works \cite{pescietc}. This idea originated --- and was studied --- in the context of qmetric, a quantum metric incorporating the \zpl\ \cite{zpl11,zpl12}. Here, I have \textit{dispensed with the notion of a qmetric}. Even the idea of \zpl\ is optional and can be eliminated in terms of Planck scale averaging.}
 It is also well known \cite{poisson} that the density of geodesics is  given by  the reciprocal of the Van-Vleck determinant $\Delta^{-1}(x,x')$. Our results suggest identifying the \mdof\ of spacetime at the event $\mathcal{P}$ as the  
the density of geodesics, with suitable coarse-graining as we approach the Planck scales. 

The expression for effective Lagrangian \eq{effL} can be evaluated if we know the exact dependence of $\ln\Delta$ on $\sigma^2$. Since we do not know this, I have used the series expansion in \eq{gexp}, which --- in turn --- leads to a Taylor series in $L_P^2$ for the terms in the effective Lagrangian. So, by construction, we get am effective Lagrangian which is made of local terms which are analytic in $L_P^2$. This, for example, ensures that in the limit of $L_P^2\to0$ we get back the classical theory with control over the corrections. In this respect, our approach is different from treating gravity as a effective field theory (see e.g., \cite{efl} for a review) which will introduce terms like $R \ln(-L^2 D^2) R$. Such non-analytic terms do not arise in our approach. (Of course, infinite order sum over a subset of terms can introduce poles which are not present at any finite order in the expansion --- like e.g., in $1+x+x^2=\cdots =(1-x)^{-1}$ developing a pole at $x=1$. But this issue can be settled only if we know the   non-perturbative dependence of $\ln\Delta$ on $\sigma^2$ (say in some specific type of geometries) so that \eq{effL} can be evaluated non-perturbatively.)

Let me now turn to several further directions of research suggested by this work.

The existence of the \zpl\ leads to an anomalous scaling dimension for the geodesic distance \cite{ad} below the scale $L_P^2$. But since the effective Lagrangian is obtained \textit{after} smearing over the scale $L_P^2$ it does not make conceptual sense to compute the anomalous scaling dimension of the geodesic distance \textit{below} the scale $L^2$. However, it has already been shown that (see \cite{2d} and references therein) the spacetime with \zpl\ behaves as though \textit{it is two-dimensional} close to Planck scales. The implications of this dimensional reduction for the geodesic distance, close to Planck scales, will be worth investigating.

Varying the effective Lagrangian, computing the quantum corrections and working out their consequences is an obvious next step. If the quantum corrections $C_{ab}$ change the left hand side of Einstein's equations to $G_{ab} + C_{ab}$, then we  would have, effectively, added matter with effective stress tensor  $-C_{ab}$. If this term violates the standard energy conditions, we have a mechanism to avoid spacetime singularities. Ascertaining what replaces the singularity  requires a non-perturbative calculation, involving the evaluation of the functional derivative $\delta \Delta/\delta g_{ab}$ without resorting to the series expansion. This is a well-defined though difficult problem.
 
Our approach has led to a clear choice for the density of \mdof:
 \begin{equation}
  f[x^i,p_i] = \Delta^{-1} [R_{ijkl} (x), L n^i]\, ; \qquad n^i n_i =1
 \end{equation}
 where $q^i=\Delta x^i\equiv Ln^i$ are Planck scale displacements in random directions. I have ignored all the dynamics contained in $n^i$ and have merely summed over it, assuming all directions are equally likely for $n^i$. 
 The next step would  be to investigate whether one can obtain  an evolution equation for $f(x^i,n^i)$ as in the case of fluid kinetics. 
 The occurrence of the factor $\ln \Delta = \ln \text{Det}\, \Delta_{ab'}$, where $\Delta_{ab'}$ is the Van Vleck matrix, suggests some possibilities. This factor will arise, for example, in a path integral over a vector field $v^a$ of the amplitude $\exp[-v^a \Delta_{ab'} v^{b'}]$. 
 All these signals the deep role played by the geodesic distance $\sigma^2 (x,x')$, Van Vleck matrix $\Delta_{ab'}$ and its determinant $\Delta$ in the \textit{microscopic} structure of spacetime.\footnote{I (and my collaborators) have been emphasizing the role of $\sigma^2$ in the microstructure of spacetime, compared to the role of metric tensor \cite{zpl11,zpl12,sigma3}, and the results here solders this point of view.} 
 
 I have worked entirely in the Euclidean space with the idea that the effective Lagrangian can be analytically continued to the Lorentzian sector. While this is a standard approach, such a transition raises  important conceptual and technical issues when we introduce $\sigma^2(x,x')$. (a) In the Euclidean sector, $\sigma^2(x,x') =0$ implies $x^i = x'^i$.  But in the Lorentzian sector the two events will coincide only when approached along a time-like or space-like geodesic; $\sigma^2(x,x') $ vanishes identically for all events connected by a null geodesic. (The limit $\sigma^2\to0$ now needs to be replaced by $\lambda\to0$ where $\lambda$ is the affine parameter. The averaging over null vectors also requires care \cite{nullav1,nullav2}.) (b) Similarly, the surface
 $\sigma^2(x,x') = \ell^2$ is a compact sphere in the Euclidean sector but is a hyperboloid in the Lorentzian sector. The limit $\ell \to 0$ will lead to a \textit{point} in the Euclidean sector but will lead to a \textit{null surface} in the Lorentzian. (This is the reason why QG effects at $\sigma^2 \lesssim L_P$ in the Euclidean sector have implications for physics close to any null surface in the Lorentzian sector.) (c) The vectors $n^i$ will become normals to null surfaces in the Lorentzian sectors changing the character of distribution function fairly drastically.
 For example, the integral of $n^an_b$ over a Euclidean sphere is finite and proportional to $\delta^a_b$; but this integral over a Lorentzian `sphere' (i.e., a hyperboloid) is divergent for non-spacelike vectors. 
 (d) It has been shown earlier \cite{tpreview1,tpreview2} that  considerations based on null surfaces lead to the following equation at leading order: $(R_{ab} - \kappa T_{ab} )\ell^a\ell^b =0$ for all null vectors $\ell^a$ in the \textit{Lorentzian} sector.  This approach, in the Lorentzian sector, also has a direct thermodynamic interpretation which is not easy to recover in the Euclidean language. 
 Thermodynamics of horizons is closely related to the existence of null surfaces which, of course, do not occur in the Euclidean sector.
 
  This connection between Lorentzian and Euclidean approaches, deserves closer scrutiny, in the light of thermodynamics of null surfaces.
 One possible approach is to work entirely in the Lorentzian sector and repeat the above analysis. The Planck scale smearing of each point in the Euclidean sector (which corresponds to smearing over a region inside a hypersphere $r^2+t_E^2\leq L_P^2$) will now be replaced by smearing of the lightcones (``thickening'' of the null surfaces) by the \zpl\ over the region between the light-cone and the hyperboloid $r^2-t^2\leq L_P^2$ (``stretched horizon'') one Planck length away. The effective Lagrangian should then arise due to  integrating out the degrees of freedom close to the null surface. The geodesic distance will be replaced by the affine distance along the null rays and should --- hopefully --- lead to similar results \cite{wip}.

The key idea in this work is a \textit{prescription} for the effective Lagrangian based on coarse-graining over Planck scales and --- like all good prescriptions --- it is well-defined, physically motivated and allows tangible calculations. All the same, it is still only a prescription and it would be interesting to derive it from a deeper layer of theory.

\section*{Acknowledgement}
I thank Sumanta Chakraborty and  Dawood Kothawala for extensive discussions and Sumanta Chakraborty for for comments on an earlier draft.
 My research  is partially supported by the J.C.Bose Fellowship of Department of Science and Technology, Government of India.


\begin{thebibliography}{000}
\bibitem{comment1} The $\Delta$ satisfies the equation $\partial_\sigma(\ln \Delta)=3\sigma^{-1}-K$ in this frame, where $K=\partial_\sigma\ln\sqrt{h}=
\partial_\sigma\ln\sqrt{\sigma^3\gamma}$ is the extrinsic curvature of the equi-geodesic surfaces. This integrates to give $\rho=\Delta^{-1}\propto\sqrt{\gamma}=
\sqrt{\gamma/\gamma_{flat}}$ with proper normalisation.
\end{thebibliography}
\end{document}